# The Phase of the Riemann Zeta Function and the Inverted Harmonic Oscillator

R.K. Bhaduri, Avinash Khare[†], and J. Law[*]

*Department of Physics and Astronomy, McMaster University,*
*Hamilton, Ontario, Canada L8S 4M1*

(June 15, 1994)

The Argand diagram is used to display some characteristics of the Riemann Zeta function. The zeros of the Zeta function on the complex plane give rise to an infinite sequence of closed loops, all passing through the origin of the diagram. This leads to the analogy with the scattering amplitude, and an approximate rule for the location of the zeros. The smooth phase of the Zeta function along the line of the zeros is related to the quantum density of states of an inverted oscillator.

The Riemann Zeta function $\zeta(s)$ of the complex variable $s = \sigma + it$ has an infinite number of zeros on the half-line $\sigma = \frac{1}{2}$ [1]. These zeros are of great interest to mathematicians from the number theoretic point of view [2,3], and to physicists interested in quantum chaos and the periodic orbit theory [4–6]. Along this line, as a function of $t$, every time $\zeta(s)$ changes sign, a discontinuous jump by $\pi$ in the phase angle is introduced. Otherwise the phase angle is a smooth function of t. The smooth part of the phase angle itself is very interesting, since it counts the number of zeros on the half-line fairly accurately. In this letter, we bring out some of the interesting properties of $\zeta(s)$ by displaying it on an Argand diagram, where $\Re\zeta(s)$ (along the x-axis), is plotted against $\Im\zeta(s)$ (y-axis). For $\sigma = 1/2$, the plot yields an infinite sequence of closed loops, one for every zero of the Zeta function, all going through the origin. This character changes drastically as one moves away from the $\sigma = 1/2$-line. The similarity in the Argand diagrams for $\zeta(1/2 + it)$ and the resonant quantum scattering amplitude is pointed out, and this analogy, although flawed, leads directly to an approximate quantization condition for the location of the zeros [7,8]. We also demonstrate that for $\sigma = 1/2$, the smooth part of the phase angle of $\zeta(s)$ is closely related to the phase shift of an inverted half harmonic oscillator.

We start by giving some standard results for the Riemann Zeta function $\zeta(s)$. Using the fundamental functional relationship between $\zeta(s)$ and $\zeta(1-s)$, it is easy to show that

$$\zeta(\frac{1}{2} - it) = (\pi)^{-it} \frac{\Gamma(\frac{1}{4} + \frac{it}{2})}{\Gamma(\frac{1}{4} - \frac{it}{2})} \zeta(\frac{1}{2} + it), \qquad (1)$$

where $\Gamma(z)$ denotes the gamma function of the argument $z$. We may further write

$$\zeta(\frac{1}{2} + it) = Z(t) \exp(-i\theta(t)), \qquad (2)$$

where $Z(t)$ is real, and $\theta(t)$ is the phase angle, with the convention that $\theta(0) = 0$. Using Eqs.(1) and (2), it follows that

$$\exp(2i\theta(t)) = \exp(-it\ln\pi) \frac{\Gamma(\frac{1}{4} + \frac{it}{2})}{\Gamma(\frac{1}{4} - \frac{it}{2})}. \qquad (3)$$

The phase $\theta$, as defined above, is smooth in the sense that it does not include the jumps by $\pi$ due to the zeros of $Z(t)$. Nevertheless, the number of zeros between 0 and $t$ on the $\sigma = 1/2$ line is counted fairly accurately by $\theta(t)$, as will become clear from the Argand diagram. Note that

$$\frac{\theta(t)}{\pi} = -\frac{t}{2\pi}\ln\pi + \frac{1}{2\pi}\Im[\ln\Gamma(\frac{1}{4} + \frac{it}{2}) - \ln\Gamma(\frac{1}{4} - \frac{it}{2})]. \qquad (4)$$

The density of zeros is given by

$$\frac{1}{\pi}\frac{d\theta}{dt} = \frac{1}{2\pi}\left[-\ln\pi + \Re[\Psi(\frac{1}{4} + i\frac{t}{2})]\right], \qquad (5)$$

where the digamma function is defined as $\Psi(z) = \Gamma'(z)/\Gamma(z)$. From the above, the asymptotic expression for $\theta(t)$ may be obtained immediately by making asymptotic expansion of the $\Gamma$ functions. We denote this by $\tilde{\theta}(t)$, and it is given by

$$\frac{1}{\pi}\tilde{\theta}(t) = (\frac{t}{2\pi})\ln(\frac{t}{2\pi}) - (\frac{t}{2\pi}) - \frac{1}{8} + \frac{1}{48\pi t} + \ldots \qquad (6)$$

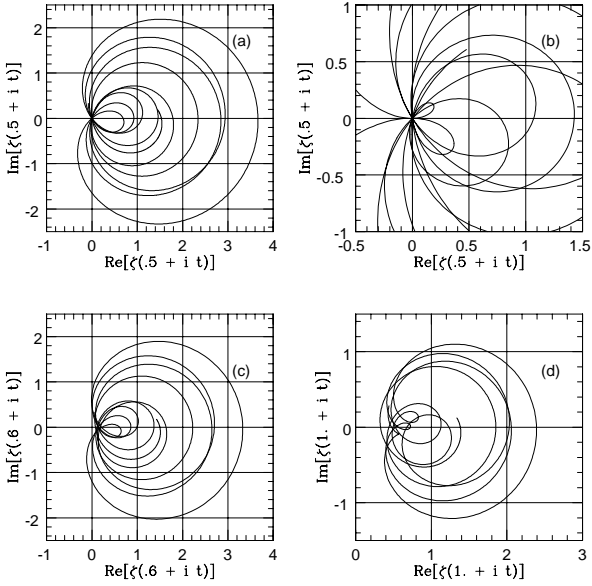

FIG. 1. The Argand diagrams for $\zeta(\sigma + it)$ for increasing $t$ and fixed $\sigma$. (a) $\zeta(1/2 + it)$ in the range $t = 9 - 50$. The lower limit for $t$ is chosen so as not to miss the Gram point flanking the lowest zero at $t = 14.13$. (b) $\zeta(1/2 + it)$ for $t = 280 - 300$, showing the two loops without the corresponding Gram points. (c) $\zeta(0.6 + it)$ for $t = 9 - 50$ to show the defocussing at the origin. (d) $\zeta(1 + it)$ for $t = 9 - 50$. Note the pronounced shift of the diagram away from the origin in this case.

To bring out some characteristics of the function $\zeta(1/2+it)$, we plot its Argand diagram in Fig.1(a) in the range $t = 9$ to $t = 50$. This shows a sequence of closed loops, one for every zero of the Zeta function. At a zero of $\zeta(1/2 + it)$, both its real and imaginary parts are zero at the same value of $t$, and therefore every loop converges at the origin. The intercepts on the real axis are the so-called "Gram points", where only the imaginary part of $\zeta(s)$ is zero, due to the phase angle $\theta(t) = n\pi$. With infrequent exceptions, there is one Gram point between two consecutive zeros of the Zeta function. The first two exceptions to this rule occur for the 126th and the 134th zeros at $t = 282.455$ and $295.584$ respectively [2]. The Argand diagram in Fig.1(b) clearly shows that for these cases, the loop structure still persists, even though the Gram points are missing. In Figs.1(c) and 1(d), Argand diagrams are drawn away from the 1/2-axis, for $\sigma = 0.6$ and $\sigma = 1$ respectively. These clearly show the defocussing at the origin due to the absence of the zeros in the Zeta function. Moreover, the number of intercepts along the real-axis in the Argand diagrams now show a large increase compared to the $\sigma = 1/2$ case, whereas the intercepts on the imaginary axis are few or nonexistant. This is a reflection of the dramatic change in the behaviour of the phase $\theta(t)$ away from the $\sigma = 1/2$ line. This is also apparent in Figs.2(a) and 2(b), where the phase angle $\theta(t)$ is plotted as a function of $t$ on and off the 1/2-axis. The abrupt change in the behaviour of the phase angle $\theta(t)$, as shown in Fig.2, is a consequence of the fact that for $\sigma \neq 1/2$, $\zeta(s)$ and $\zeta(1-s)$ are not complex conjugate of each other. Therefore Eq.(3) no longer holds in such a situation. The "chaotic" characteristic of the phase angle off the $1/2-$ axis, is, of course, well-known.

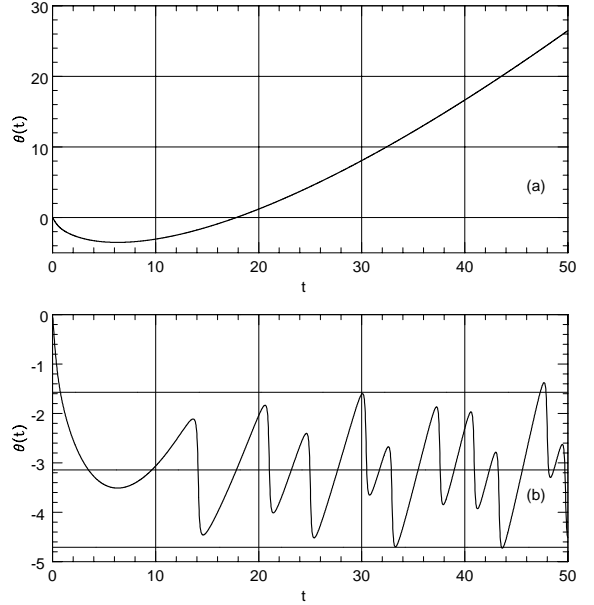

FIG. 2. The exact phase angle of $\zeta(\sigma + it)$ is plotted as a function of $t = 0 - 50$ for (a) $\sigma = 1/2$ and (b) $\sigma = 0.6$. See text for discussion.

Indeed, the phase angle for $s = 1 + it$ is well studied, and is related to the quantum scattering phase shift of a particle on a surface of constant negative curvature [13–15].

Finally, in Figs. 3(a) and 3(b), the Argand diagrams of the Zeta function are drawn for a much larger range of $t$, from 1 to 500, on and off the $1/2-$axis. Note that the scale for $\sigma = 1$ is expanded compared to that for $\sigma = 1/2$. Borrowing from the terminology of the motion of a particle in phase space, it is as if there is an "attractor" at the origin for $\sigma = 1/2$ (Fig. 3(a)), which is absent from the more disorderly tracks of Fig. 3(b), which is drawn along the $\sigma = 1$ line. The latter figure also shows that the real part of the Zeta function is always positive for $\sigma = 1$ for this entire range of $t$.

$$f_l(k) = (\eta_l \, exp(2i\delta_l) - 1)/2ik. \qquad (7)$$

Here $l$ refers to the angular momentum, and $k$ the wave number. Note that $\Im f_l(k)$ is never negative, since the inelasticity parameter $\eta_l$ is always less than one. One generally plots an Argand diagram with $2k\Im f_l(k)$ along the $y-$axis and $2k\Re f_l(k)$ along the x-axis for various values of $k$. For the case of no inelasticity ($\eta_l = 1$) and a single $\Phi$ resonance, the Argand diagram is a perfect circle with unit radius, with the center on the imaginary axis at 1. By comparing this with Fig.1(a) at $\sigma = 1/2$, we see that the real and the imaginary parts are interchanged in the latter, but otherwise there is a strong similarity, with many of the loops having inelasticity. This analogy is flawed, however, since $\Re\zeta(1/2 + it)$ does become negative in small islands of $t$. Nevertheless, if these islands are ignored, then the phase shift $\delta_l$ may be identified with the phase angle $\theta + \pi/2$, with each closed loop in Fig.1(a) being regarded as in isolated resonance. In this approximation, the Gram points occur as before for $\sin\theta = 0$, while the zeros of $\zeta(1/2 + it)$ are given by the condition

$$\cos\theta = 0, \quad \theta = (m + 1/2)\pi, \quad m = 1, 2, \ldots . \qquad (8)$$

This condition for the location of the zeros was also obtained by Berry [7] from the first term in his approximate formula. Eq.(8) has roots that yield the zeros on the $1/2-$axis with an error of at most 3 percent.

As mentioned earlier, for $\sigma = 1$, the phase $\theta(t)$ is intimately connected to the quantum scattering of a particle on a saddle-like surface. On the $\sigma = 1/2$ line, our analogy with the scattering amplitude also suggests that the phase angle $\theta(t)$ is related to a scattering phase shift. We now demonstrate that the scattering of a nonrelativistic particle by an inverted harmonic oscillator with a hard wall at the origin generates a phase shift that is closely related to $\theta(t)$. Indeed, we show that the quantum density of states for this problem is essentially the same as Eq.(5) for the density of the zeros. Consider the Schrödinger equation for $x \geq 0$,

$$-\frac{\hbar^2}{2m}\frac{d^2}{dx^2}\Phi - \frac{1}{2}m\omega^2 x^2 \Phi = E\Phi, \qquad (9)$$

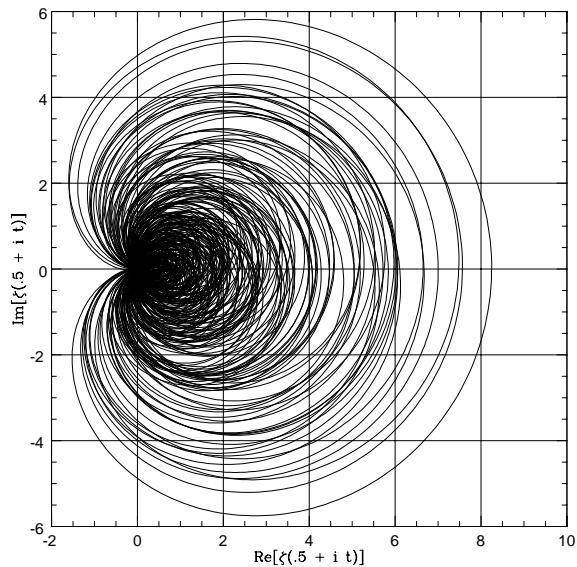

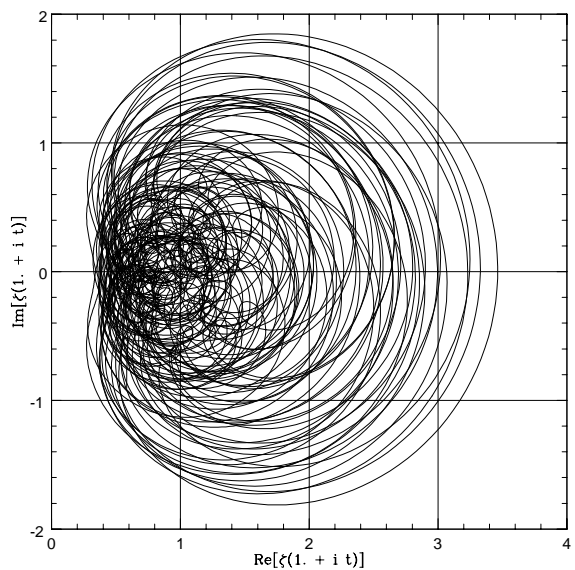

FIG. 3. The Argand diagrams of the Riemann Zeta function for the wider range of $t-$values, from $t = 9 - 500$. (a) $\zeta(1/2 + it)$, and (b) $\zeta(1 + it)$.

The loop structure of the Zeta function at $\sigma = 1/2$, with some near-circular shapes, is reminiscent of the Argand plots for the scattering amplitudes of different partial waves in the analysis of resonances, for example in pion-nucleon scattering [9]. Consider the partial wave amplitude $f_l(k)$, defined in terms of the partial wave phase-shift $\delta_l(k)$ and the inelaticity parameter $\eta_l(k)$, and impose the boundary condition that the wave function $\Phi$ vanishes at the origin. Putting $x^2 = y$, $\Phi = y^{-\frac{1}{4}}\phi$, it becomes

$$\frac{d^2}{dy^2}\phi - \frac{l(l+1)}{y^2}\phi - \frac{kt}{y}\phi + k^2\phi = 0. \qquad (10)$$

In the above equation,

$$l = -\frac{1}{4}, \quad k = \frac{m\omega}{2\hbar}, \quad t = -\frac{E}{\hbar\omega}. \qquad (11)$$

This is effectively a three dimensional Schrödinger equation for a repulsive Coulomb potential in the variable $y$.

To obtain the phase shift, we write the asymptotic solution of the above equation [16] as

$$\phi(y) \sim \sin(ky - \frac{t}{2}\ln(2ky) - \frac{l\pi}{2} + \eta_l), \quad (12)$$

where $\eta_l$ is the phase shift with respect to the distorted Coulomb wave, given by $\arg \Gamma(l+1+it/2)$. For our one-dimensional problem, only $l = -1/4$ is relevant. For this case, omitting the subscript $l$, the phase shift $\eta$ is

$$\eta = \arg \Gamma(\frac{3}{4} + \frac{it}{2}). \quad (13)$$

Using the identity [10]

$$\Gamma(\frac{1}{4} + iy)\Gamma(\frac{3}{4} - iy) = \frac{\pi\sqrt{2}}{\cosh \pi y + i \sinh \pi y}. \quad (14)$$

the number of quantum states $n(t)$, between 0 and $t$, is then given by

$$n(t) = \frac{\eta(t)}{\pi} = \frac{C}{2\pi} + \frac{1}{2\pi}\Im\left[\ln \Gamma(\frac{1}{4} + i\frac{t}{2}) - \ln \Gamma(\frac{1}{4} - i\frac{t}{2})\right]. \quad (15)$$

In the above equation, $C$ is a smooth function given by

$$C = \frac{\pi}{2} - \tan^{-1}(\operatorname{cosech}\pi t). \quad (16)$$

Note from above that the expression for $\eta(t)$ is not quite identical to $\theta(t)$ as defined in Eq.(4). However, their derivatives, the quantum density of states, only differ by a constant and an exponentially small term. It should also be pointed out that even if we had started with a full inverted harmonic oscillator (rather than the half-oscillator), the same conclusion would be reached, even though there may arise some nonuniqueness in the choice of the boundary condition. The inverted harmonic oscillator problem was studied by a number of authors in the past [11,12] in relation to time-delay, and by others [17-20] in connection to string theory. No connection, however, was made to the phase of $\zeta(1/2 + it)$.

In summary, we recapitulate the main points made in this paper. Traditionally, the behaviour of the zeros of the Riemann Zeta function on the $1/2-$axis is associated with the bound state problem of a quantum Hamiltonian. We, on the other hand, focus on the scattering problem. We use the Argand diagram construction for the Riemann Zeta function to visualize how the smooth phase $\theta(t)$ acts as a counter for the zeros of $Z(t)$. This leads to the analogy with the scattering amplitude, and the approximate condition ( Eq.(8) ) for the location of the zeros. We also find a connection between the phase angle of the Zeta function on the $\sigma = 1/2$ line and the quantum phase shift of a potential. It is perhaps not surprising that this potential is as simple as the inverted harmonic oscillator, and not something more complicated, as is the case for the chaotic phase for $\sigma = 1$ [13-15]. Finally, we suggest that the Argand diagram construction may also be useful for the Selberg Zeta function, and shed light on the quantisation condition for quantum chaos [8].

One of the authors (A.K.) would like to acknowledge the hospitality of the physics department of McMaster University. This research was supported by grants from NSERC of Canada.